\documentclass[aps,prb,twocolumn,showpacs,amsmath]{revtex4}
\usepackage{latexsym}
\usepackage{amssymb}
\usepackage{bm}% bold math
\usepackage{graphicx}

\newcommand{\pdag}{{\phantom{\dagger}}}
\newcommand{\bq}{\begin{equation}}
\newcommand{\eq}{\end{equation}}
\newcommand{\bn}{\begin{eqnarray}}
\newcommand{\en}{\end{eqnarray}}

\begin{document}

\title{Finite-frequency current (shot) noise in coherent resonant tunneling through a %%@
coupled-quantum-dot interferometer}

\author{Bing Dong} 
\affiliation{Department of Physics, Shanghai Jiaotong University,
1954 Huashan Road, Shanghai 200030, China}

\author{X. L. Lei} 
\affiliation{Department of Physics, Shanghai Jiaotong University,
1954 Huashan Road, Shanghai 200030, China}

\author{N. J. M. Horing}
\affiliation{Department of Physics and Engineering Physics, Stevens Institute of Technology, %%@
Hoboken, New Jersey 07030, USA}

\begin{abstract}

We examine the shot noise spectrum properties of coherent resonant tunneling in coupled quantum %%@
dots in both series and parallel arrangements by means of quantum rate equations and MacDonald's %%@
formula.
Our results show that, for a series-CQD with a relatively high dot-dot hopping $\Omega$, %%@
$\Omega/\Gamma\gtrsim 1$ ($\Gamma$ denotes the dot-lead tunnel-coupling strength), the noise %%@
spectrum exhibits a dip at the Rabi frequency, $2\Omega$, in the case of noninteracting %%@
electrons, but the dip is supplanted by a peak in the case of strong Coulomb repulsion; %%@
furthermore, it becomes a dip again for a completely symmetric parallel-CQD by tuning enclosed %%@
magnetic-flux.   
 
\end{abstract}


\maketitle

\section{Introduction}

The study of current noise through a quantum dot (QD) has recently become an emerging topic of %%@
interest in mesoscopic physics, because measurements of shot noise can yield more information %%@
about the microscopic mechanisms of transport than can conductance measurements %%@
alone.\cite{Blanter,Beenakker} In particular, finite-frequency noise provides direct access to %%@
the intrinsic dynamics of a mesoscopic system. 
For example, it has been reported that coherent intrinsic Rabi oscillations in coupled QDs (CQD) %%@
causes a dip at the Rabi frequency in the noise spectrum.\cite{Sun,Aguado,Djuric,Djuric1} On the %%@
contrary, a peak at the Zeeman frequency in the noise spectrum has been addressed for the %%@
coherent magnetotransport through an interacting QD with Zeeman-splitting levels.\cite{Fedichkin} %%@
Later, the authors have analyzed further the role of Coulomb interaction in the magnetotransport %%@
through a QD and pointed out that the noise spectrum indeed exhibits a peak at the Zeeman %%@
frequency in the regime of Coulomb blockade, but it becomes a dip for noninteracting %%@
electrons.\cite{Gurvitz3} So it is natural to ask what will happen in the noise spectrum of a CQD %%@
if the interdot Coulomb interaction is taken into account. This constitutes the first purpose of %%@
the present paper.    

On the other hand, coherent tunneling through a CQD in a parallel arrangement between two %%@
electrodes is quite interesting and has attracted numerous investigations because both interdot %%@
coherence and also quantum interference between two distinct magnetic-flux pathways play an %%@
important role in determining the transport properties of this %%@
system.\cite{Holleitner1,Holleitner2,ChenJ,Shahbazyan,Loss,Konig,Kubala,Ladron%%@
,Dong1} The interference pattern can be changed from destructive to constructive by tuning the %%@
magnetic-flux piercing the enclosed two pathways. Even though it has been reported that the %%@
interference effect produces a peak or a dip in the noise spectrum of a QD with two energy %%@
levels, depending on the relative phase of the two levels carrying the current,\cite{GurvitzIEEE} %%@
yet it is not quite clear about the combination of the interdot Coulomb repulsion and %%@
interference effect, which is the second purpose of our studies.
 
In this paper, we analyze {\em finite-frequency} shot noise of first-order coherent tunneling %%@
through a CQD in both series and parallel configurations using quantum rate equations (QREs) and %%@
MacDonald's formula.
In Sec.~II, we introduce the model Hamiltonian of the system under investigation. In the %%@
following two sections, we consider  the two cases of noninteracting electrons and interacting %%@
electrons respectively. In these sections, we discuss the respective number-resolved QREs for %%@
describing first-order resonant tunneling at extremely high bias-voltage, which can be obtained %%@
from our previous derivation.\cite{Dong1,Dong4} Then we apply MacDonald's scheme to calculate the %%@
tunneling current and its noise spectrum.\cite{Gurvitz3,MacDonald} In particular, we derive %%@
explicit analytical expressions for the noise spectrum and charge fluctuation spectrum of a %%@
series-CQD and a completely symmetric parallel-CQD. In Sec.~V, we present numerical calculations %%@
and point out how strong Coulomb repulsion and quantum interference influence the noise spectrum. %%@
We find that the noise spectrum exhibits a dip at the Rabi frequency for a noninteracting %%@
series-CQD with large dot-dot hopping strength, but the dip is supplanted by a peak in the case %%@
of strong Coulomb repulsion. As the coupling strength of the additional path branch is increased, %%@
the Coulomb interaction-induced peak in the noise spectrum gradually becomes a Fano-type peak, %%@
and finally emerges as a dip in the case of nonzero enclosed magnetic-flux. Sec.~VI summaries our %%@
conclusions.

\section{Model Hamiltonian}

The Hamiltonian of the parallel-CQD interferometer connected to two normal leads is written as  
\bq
H=H_{L}+H_{R}+H_{D}+H_{T}, \label{ham}
\eq
with     
\bn
H_{\eta} &=& \sum_{{\bf k}} \varepsilon_{\eta{\bf k}} a_{\eta{\bf k}}^\dagger a_{\eta{\bf %%@
k}}^\pdag, \\
H_{D} &=& \sum_{j=1,2} \varepsilon_{j}c_{j}^\dagger c_{j}^\pdag + U c_{1}^\dagger c_{1}^\pdag %%@
c_{2}^\dagger c_{2}^\pdag + \Omega ( c_{1}^\dagger c_{2}^\pdag + c_{2}^\dagger c_{1}^\pdag), %%@
\label{Hqd} \\
H_{T} &=& \sum_{{\bf k}} [(V_{L1} e^{i\varphi /4} a_{L {\bf k} }^{\dagger } + V_{R1} e^{-i\varphi %%@
/4} a_{R {\bf k} }^{\dagger } ) c_{1} \cr
&& + (V_{L2} e^{-i\varphi /4} a_{L {\bf k} }^{\dagger } + V_{R2} e^{i\varphi /4} a_{R {\bf k} %%@
}^{\dagger } ) c_{2} +{\rm {H.c.}}],  \label{hamiltonianT}
\en
where $a_{\eta{\bf k}}^\dagger$ ($a_{\eta{\bf k}}$) is the creation (annihilation) operator of an %%@
electron with momentum ${\bf k}$, and energy $\varepsilon_{\eta {\bf k}}$ in lead $\eta$($=L,R$), %%@
and $c_{j}^\dagger$ ($c_{j}$) is the creation (annihilation) operator for a spinless electron %%@
with energy $\varepsilon_j$ in the $j$th QD ($j=1,2$). $\varepsilon_{1(2)}=\varepsilon_d \pm %%@
\delta$, with $\delta$ being the bare mismatch between the two bare levels. $U$ represents the %%@
interdot Coulomb interaction, and $\Omega$ denotes hopping strength between the two QDs. 
$V_{\eta j}$ represents the tunneling matrix element between lead $\eta$ and dot $j$, and is %%@
assumed to be real and independent of energy. In this paper, we assume $V_{L1}=V_{R2}$ and %%@
$V_{L2}=V_{R1}$. The factor $e^{\pm i\varphi /4}$ is the accumulated Peierls phase due to the %%@
magnetic-flux $\Phi$ ($\varphi\equiv 2\pi \Phi/\Phi_0$, $\Phi_0\equiv hc/e$ is the magnetic-flux %%@
quantum) which penetrates the area enclosed by the two tunneling pathways of the Aharonov-Bohm %%@
(AB) interferometer. Such a QD AB interferometer has been realized in recent %%@
experiments.\cite{Holleitner1,Holleitner2,ChenJ} Furthermore, we assume the two electrodes to be %%@
Markovian stationary reservoirs with a flat density of states $\varrho$.

In this coupled-QD (CQD) system, there are a total of four possible states for the present %%@
system: (1) the whole system is empty, $|0\rangle_{1} |0\rangle_{2} $, and its energy is zero; %%@
(2) the first QD is singly occupied by an electron, $|1 \rangle_{1} |0\rangle_{2}$, and its %%@
energy is $\varepsilon_{1}$; (3) the second QD is singly occupied, $|0\rangle_{1} |1\rangle_{2}$, %%@
and its energy is $\varepsilon_2$; and (4) both QDs are occupied, $|1\rangle_{1} |1\rangle_{2}$, %%@
and its energy is $\varepsilon_1+\varepsilon_2+U$. Of course, if the interdot Coulomb repulsion %%@
is assumed to be infinite, the double-occupation is prohibited. Therefore, as far as the four %%@
possible single electronic states are considered as the basis, the density-matrix elements are %%@
expressed as $\hat\rho_{00}=|0\rangle_{1} |0 \rangle_{2} {}_{2}\langle 0| {}_{1}\langle 0|$, %%@
$\hat\rho_{11}=|1\rangle_{1} |0 \rangle_{2} {}_{2}\langle 0| {}_{1}\langle 1|$, %%@
$\hat\rho_{22}=|0\rangle_{1} |1 \rangle_{2} {}_{2}\langle 1 | {}_{1}\langle 0|$, %%@
$\hat\rho_{dd}=|1\rangle_{1} |1 \rangle_{2} {}_{2}\langle 1 | {}_{1}\langle 1|$, and %%@
$\hat\rho_{12}=|1\rangle_{1} |0 \rangle_{2} {}_{2}\langle 1| {}_{1}\langle 0|$. The statistical %%@
expectation values of the diagonal elements of the density-matrix, $\rho_{00}=\langle %%@
\hat\rho_{00}\rangle$, $\rho_{jj}=\langle \hat\rho_{jj}\rangle$ ($j=1,2$), and $\rho_{dd}=\langle %%@
\hat\rho_{dd}\rangle$ describe the occupation probabilities of the electronic levels for the %%@
system being empty, singly occupied in the $j$th QD by an electron, or doubly occupied in both %%@
QDs, respectively. The off-diagonal term $\rho_{12}=\langle \hat\rho_{12}\rangle$ describes the %%@
coherent superposition of two electronic single occupation states, $|1 \rangle_{1} |0\rangle_{2}$ %%@
and $|0\rangle_{1} |1\rangle_{2}$. 

In the limit of sufficiently large bias-voltage [$V\gg \Gamma,\Gamma'$ (the tunnel-coupling %%@
strengths between dots and leads, which are defined below), and $\Omega$], electronic tunneling %%@
through this system in the first-order approximation can be described by recently derived QREs %%@
for the dynamic evolution of the reduced density matrix
elements of the CQD, $\rho_{ab}(t)$ ($a,b=\{0, 1, 2, %%@
d\}$).\cite{Dong1,GurvitzIEEE,Dong4,Stoof,Gurvitz,Gurvitz1,Dong} In the following, we employ %%@
MacDonald's formula, in conjunction with the number-resolved QREs, to evaluate the noise spectrum %%@
of coherent resonant tunneling in a CQD.

\section{CQD in the case of extremely large bias-voltage and no interdot Coulomb repulsion}

First, we consider the absence of interdot Coulomb repulsion, $U=0$, in the case of extremely %%@
large bias-voltage and zero temperature. Throughout, we will use units with $\hbar=k_B=e=1$.    

\subsection{Number-resolved QREs}

In the discussion of frequency-dependent current noise in the CQD interferometer, we employ %%@
MacDonald's formula for shot noise\cite{MacDonald} based on the number-resolved version of the %%@
QREs, in accordance with the number of achieved tunneling events.\cite{Chen} 
We introduce the number-resolved density matrices $\rho_{ab}^{(n)}(t)$ ($a,b=\{0,1,2,d\}$), %%@
representing the probability that the system is in the electronic state $|a\rangle$ ($a=b$), or %%@
in the quantum superposition state ($a\neq b$), at time $t$ together with $n$ electrons entering %%@
into the left lead due to tunneling events. Obviously, $\rho_{ab}(t)=\sum_{n} \rho_{ab}^{(n)}(t)$ %%@
and the associated number-resolved QREs in the case of extremely large bias-voltage and zero %%@
temperature are:
\begin{subequations}
\label{rateqU0} 
\bn
\dot{\rho}_{00}^{(n)}&=& \Gamma' \rho_{11}^{(n-1)} + \Gamma \rho_{22}^{(n-1)} - (\Gamma + %%@
\Gamma') \rho_{00}^{(n)} \cr
&& + \sqrt{\Gamma\Gamma'} (e^{-i\varphi/2} \rho_{12}^{(n-1)} + e^{i\varphi/2} \rho_{21}^{(n-1)}), %%@
\label{nrc01} \\
\dot{\rho}_{11}^{(n)} &=& \Gamma \rho_{00}^{(n)} - 2 \Gamma' \rho_{11}^{(n)} +\Gamma %%@
\rho_{dd}^{(n-1)} + i \Omega ( \rho_{12}^{(n)}-\rho_{21}^{(n)}) \cr
&& -\frac{1}{2} \sqrt{\Gamma\Gamma'} (e^{-i\varphi/2} - e^{i\varphi/2}) (\rho_{12}^{(n)}- %%@
\rho_{21}^{(n)}), \label{nrc11} \\
\dot{\rho}_{22}^{(n)} &=& \Gamma' \rho_{00}^{(n)} - 2\Gamma \rho_{22}^{(n)} + \Gamma' %%@
\rho_{dd}^{(n-1)} -i\Omega (\rho_{12}^{(n)}-\rho_{21}^{(n)}) \cr
&& -\frac{1}{2} \sqrt{\Gamma\Gamma'} (e^{-i\varphi/2} - e^{i\varphi/2}) (\rho_{12}^{(n)}- %%@
\rho_{21}^{(n)}), \label{nrc21} \\
\dot{\rho}_{dd}^{(n)} &=& \Gamma' \rho_{11}^{(n)} + \Gamma \rho_{22}^{(n)} - (\Gamma+ \Gamma') %%@
\rho_{dd}^{(n)} \cr
&& - \sqrt{\Gamma\Gamma'}e^{i\varphi /2} \rho_{12}^{(n)} - \sqrt{\Gamma\Gamma'}e^{-i\varphi /2} %%@
\rho_{21}^{(n)}, \\
\dot{\rho}_{12}^{(n)} &=& i\Omega (\rho_{11}^{(n)}-\rho_{22}^{(n)}) - (\Gamma + \Gamma') %%@
\rho_{12}^{(n)} \cr
&& + \sqrt{\Gamma\Gamma'} e^{-i\varphi/2} \rho_{00}^{(n)} - \sqrt{\Gamma\Gamma'} e^{i\varphi/2} %%@
\rho_{dd}^{(n-1)} \cr
&& - \frac{1}{2} \sqrt{\Gamma\Gamma'} (e^{i\varphi/2}- e^{-i\varphi/2} ) (\rho_{11}^{(n)} + %%@
\rho_{22}^{(n)}), \label{nrc31} 
\en
\end{subequations}
together with the normalization relation $\rho_{00}+\rho_{11}+\rho_{22} + \rho_{dd}=1$. The %%@
adjoint equation of Eq.~(\ref{nrc31}) gives
the equation of motion for $\rho_{21}^{(n)}$. The constant parameters $\Gamma=2\pi \varrho_{L} %%@
|V_{L1(R2)}|^2$ and $\Gamma'=2\pi \varrho_{L} |V_{L2(R1)}|^2$ represent the strengths of %%@
tunnel-couplings between dots and leads;   
$\sqrt{\Gamma\Gamma'}$ describes the interference in tunneling events arising from different %%@
pathways. From these number-resolved QREs, we can readily deduce the usual QREs for the reduced %%@
density matrix elements, ${\bm \rho}(t) = (\rho_{00}, \rho_{11}, \rho_{22}, \rho_{dd}, \rho_{12}, %%@
\rho_{21})^{T}$, as: 
\bq
\dot {\bm \rho}(t) = {\cal M}{\bm \rho}(t), \label{qre}
\eq
where ${\cal M}$ can be easily read from Eqs.~(\ref{nrc01})-(\ref{nrc31}) as:
\begin{widetext}
\bq
{\cal M}= \left ( 
\begin{array}{cccccc}
-(\Gamma+\Gamma') & \Gamma' & \Gamma & 0 & \sqrt{\Gamma\Gamma'}e^{-i\varphi/2} & %%@
\sqrt{\Gamma\Gamma'}e^{i\varphi/2} \\
\Gamma & -2\Gamma' & 0 & \Gamma & -i\Omega-\frac{\sqrt{\Gamma\Gamma'}}{2}e^{-i\varphi/2} & %%@
i\Omega-\frac{\sqrt{\Gamma\Gamma'}}{2} e^{i\varphi/2} \\
\Gamma' & 0 & -2\Gamma & \Gamma' & i\Omega-\frac{\sqrt{\Gamma\Gamma'}}{2} e^{-i\varphi/2} & %%@
-i\Omega-\frac{\sqrt{\Gamma\Gamma'}}{2} e^{i\varphi/2} \\
0 & \Gamma' & \Gamma & -(\Gamma+\Gamma') & -\sqrt{\Gamma\Gamma'}e^{i\varphi/2} & %%@
-\sqrt{\Gamma\Gamma'}e^{-i\varphi/2} \\ 
\sqrt{\Gamma\Gamma'} e^{-i\varphi/2} & i\Omega-i\sqrt{\Gamma\Gamma'}\sin(\varphi/2) & %%@
-i\Omega-i\sqrt{\Gamma\Gamma'}\sin(\varphi/2) & -\sqrt{\Gamma\Gamma'}e^{i\varphi/2} & %%@
-\frac{1}{2}(\Gamma+\Gamma') & 0 \\
\sqrt{\Gamma\Gamma'} e^{i\varphi/2} & -i\Omega+i\sqrt{\Gamma\Gamma'}\sin(\varphi/2) & %%@
i\Omega+i\sqrt{\Gamma\Gamma'}\sin(\varphi/2) & -\sqrt{\Gamma\Gamma'}e^{-i\varphi/2} & 0 & %%@
-\frac{1}{2}(\Gamma+\Gamma') 
\end{array}
\right ),
\eq
\end{widetext}
  
The current flowing through the system can be evaluated as the time rate of change of electron %%@
number in the right lead 
\bq
I_{R}(t) = \dot N_{R}(t) =\frac{d}{dt} \sum_{n} n P^{(n)}(t), \label{Inr1}
\eq
where
\bq
P^{(n)}(t) = \rho_{00}^{(n)}(t)+ \rho_{11}^{(n)}(t)+ \rho_{22}^{(n)}(t)+ \rho_{dd}^{(n)}(t) 
\eq
is the total probability of transferring $n$ electrons into the right lead by time $t$. From the %%@
number-resolved QREs, Eqs.~(\ref{nrc01})--(\ref{nrc31}), it is readily shown that $I_R(t)$ is %%@
given by:
\bn
I_{R}(t)&=&\Gamma' \rho_{11}+ \Gamma \rho_{22}+ (\Gamma+\Gamma') \rho_{dd} \cr
&&+ \sqrt{\Gamma\Gamma'} (e^{-i\varphi/2} \rho_{12} + e^{i\varphi/2} \rho_{21}). \label{Inr2}
\en
Substituting the stationary solution for ${\bm \rho}_{st}$ of Eq.~(\ref{qre}) into %%@
Eq.~(\ref{Inr2}), we can evaluate the stationary current $I$. In particular, we obtain analytical %%@
expressions for several magnetic-fluxes, $\varphi$: ($x=\Omega/\Gamma$ and %%@
$\gamma=\Gamma'/\Gamma$)  
\bq
I = \left \{
\begin{array}{ccl}
\frac{2(\gamma+1)(\gamma+x^2)}{(\gamma+1)^2 + 4x^2} \Gamma, & \, & \text {$\varphi=0$ and $2\pi$} %%@
\\
\frac{2(\gamma+1) x^2}{(\gamma+1)^2 + 4x^2} \Gamma, & \, & \varphi=\pm \pi.
\end{array}
\right.
\eq
If $\Gamma'=0$ ($\gamma=0$), the system reduces to a series-CQD, and, correspondingly, the %%@
stationary current becomes
\bq
I=\frac{2x^2}{1 + 4x^2} \Gamma. \label{eq:current-of-series}
\eq
On the other hand, for a completely symmetrical parallel-CQD, $\Gamma'=\Gamma$ ($\gamma=1$), we %%@
have
\bq
I = \left \{
\begin{array}{ccl}
\Gamma, & \, & \text {$\varphi=0$ and $2\pi$} \\
\frac{x^2}{1 + x^2} \Gamma, & \, & \varphi=\pm \pi.
\end{array}
\right. \label{eq:current-parallel-CQDU0}
\eq

\subsection{MacDonald's formula for shot noise spectrum}

The frequency-dependent shot noise spectrum with respect to the right lead can be defined in %%@
terms of $P^{(n)}(t)$ using the standard technique based on MacDonald's formula:\cite{MacDonald}
\bq
S_{R}(\omega)=2\omega\int_0^\infty dt \sin(\omega t) \left [ \frac{d}{dt} \sum_{n} n^2 P^{(n)}(t) %%@
- 2I^2 t \right ]. \label{snnr}
\eq
In this formulation, the long-time linear behavior of the first term inside the square brackets %%@
on the right hand side is canceled by the latter term, $2I^2 t$. Then the noise spectrum can be %%@
evaluated in the form
\bq
S_R(\omega)= i\omega [ {\cal P}(i\omega)- {\cal P}(-i\omega)], \label{snnr2}
\eq
with 
\bq
{\cal P}(s)=\int_0^\infty dt \, e^{-s t} \frac{d}{dt} \sum_{n} n^2 P^{(n)}(t).
\eq
Here, the stationary-current-related term of Eq.~(\ref{snnr}) gives no contribution.

To evaluate $S_{R}(\omega)$, we define an auxiliary function $G_{ab}(t)$ as
\bq
G_{ab}(t)=\sum_{n} n \rho_{ab}^{(n)}(t),
\eq
whose equations of motion can be readily deduced employing the number-resolved QREs, %%@
Eqs~(\ref{nrc01})--(\ref{nrc31}), in matrix form:
\bq
\dot{\bm G}(t)={\cal M}{\bm G}(t) + {\cal G}{\bm \rho}(t), \label{qre:g}
\eq
with ${\bm G}(t) = (G_{00},G_{11},G_{22},G_{dd},G_{12},G_{21})^{T}$ and
\bq
{\cal G}=\left (
\begin{array}{cccccc}
0 & \Gamma' & \Gamma & 0 & \sqrt{\Gamma\Gamma'}e^{-i\varphi/2} & \sqrt{\Gamma\Gamma'} %%@
e^{i\varphi/2} \\
0 & 0 & 0 & \Gamma & 0 & 0 \\ 
0 & 0 & 0 & \Gamma' & 0 & 0 \\ 
0 & 0 & 0 & 0 & 0 & 0 \\ 
0 & 0 & 0 & -\sqrt{\Gamma\Gamma'}e^{i\varphi/2} & 0 & 0 \\
0 & 0 & 0 & -\sqrt{\Gamma\Gamma'}e^{-i\varphi/2} & 0 & 0
\end{array} 
\right ).
\eq 
With the help of the number-resolved QREs Eqs.~(\ref{nrc01})--(\ref{nrc31}), we find
\bn
{\cal P}(s) &=& \bigl ( \Gamma' [2G_{11}(s) + \rho_{11}(s)] + \Gamma [2G_{22}(s) + \rho_{22}(s)] %%@
\cr
&& + (\Gamma+\Gamma') [2G_{dd}(s)+\rho_{dd}(s)] \cr 
&& + \sqrt{\Gamma\Gamma'} \left \{ e^{-i\varphi/2} [2G_{12}(s)+ \rho_{12}(s)] \right. \cr
&& \left. + e^{i\varphi/2} [2G_{21}(s)+ \rho_{21}(s)] \right \} \bigr ), \label{noiseq}
\en 
where $G_{ab}(s)$ and $\rho_{ab}(s)$ are the Laplace transforms of $G_{ab}(t)$ and %%@
$\rho_{ab}(t)$, respectively:
\bq
G(\rho)_{ab}(s)=\int_0^\infty dt \, e^{-st} G(\rho)_{ab}(t).
\eq
Accordingly, we can readily evaluate ${\bm \rho}(s)$ by Laplace transforming its equation of %%@
motion, Eq.~(\ref{qre}), with the initial condition, ${\bm \rho}(0)={\bm \rho}_{st}$:
\bq
{\bm \rho}(s)=(s{\bm I}- {\cal M})^{-1} {\bm \rho}_{st}.
\eq
Thus, applying the Laplace transform to Eq.~(\ref{qre:g}) yields
\bq
{\bm G}(s) = (s {\bm I}-{\cal M})^{-1} {\cal G} {\bm \rho}(s).
\eq
Substituting the Laplace transform solutions for ${\bm G}(s)$ and ${\bm \rho}(s)$ into %%@
Eqs.~(\ref{noiseq}) and (\ref{snnr2}), we can evaluate the noise spectrum.
It should be noted that due to the inherent long-time stability of the physical system under %%@
consideration, all real parts of the nonzero poles of ${\bm \rho}(s)$ and ${\bm G}(s)$ are %%@
negative definite. Consequently,  the divergent terms of the partial fraction expansions of ${\bm %%@
\rho}(s)$ and ${\bm G}(s)$ at $s\rightarrow 0$ entirely determine the long-time behavior of the %%@
auxiliary functions, i.e., the zero-frequency shot noise, Eq.~(\ref{snnr}) at %%@
$\omega=0$.\cite{Dong4} 

Furthermore, according to the Ramo-Shockley theorem,\cite{Ramo,Shockley} the average current %%@
$I_c$ is given by
\bq
I_c(t)=\alpha I_L(t)+ \beta I_R(t).
\eq
Here the coefficients $\alpha$ and $\beta$, which satisfy $\alpha+ \beta=1$, depend on the %%@
barrier geometry. For simplicity we take $\alpha= \beta= \frac{1}{2}$. Considering charge %%@
conservation, $I_L=I_R + \dot Q$, with \bq
Q(t)=\sum_n [\rho_{11}^{(n)}(t) + \rho_{22}^{(n)}(t) + 2\rho_{dd}^{(n)}(t)]
\eq
as the total charge in the CQD,\cite{Korotkov} we have
\bq
I_c(t)I_c(0)=\alpha I_L(t) I_L(0) + \beta I_R(t) I_R(0) - \alpha\beta \dot Q(t) \dot Q(0).
\eq
Thus, the total noise spectrum is given by
\bq
S(\omega)=\alpha S_L(\omega) + \beta S_R(\omega) - \alpha\beta \omega^2 S_Q(\omega),
\eq
where $S_L(\omega)$ is the noise spectrum with regard to the left lead and $S_Q(\omega)$ is the %%@
Fourier transform of the symmetrical charge correlation function. Generally, one can show that %%@
$S_{L}(\omega)=S_R(\omega)$ and calculate $S_Q(\omega)$ as\cite{Gurvitz3}
\bq
S_Q(\omega)= 2 [{\cal Q}(i\omega)+ {\cal Q}(-i\omega)],
\eq
with
\bn
{\cal Q}(s) &=& \int_0^\infty dt \, e^{-s t} Q(t) \cr
&=& \rho_{11}'(s)+ \rho_{22}'(s) + 2\rho_{dd}'(s).
\en
Likewise, we evaluate the Laplace transform solution for ${\bm \rho}'(s) = (\rho_{00}', %%@
\rho_{11}', \rho_{22}', \rho_{dd}', \rho_{12}', \rho_{21}')^{T}$ with the associated initial %%@
condition, ${\bm \rho}'(0)=(0, \rho_{11}^{st}, \rho_{22}^{st}, \rho_{dd}^{st}, 0, 0)^{T}$, %%@
through
\bq
{\bm \rho}'(s)=(s{\bm I}- {\cal M})^{-1} {\bm \rho}'(0).
\eq

Straightforwardly, we obtain analytical expressions for the frequency-dependent noise for several %%@
special cases. When $\Gamma'=0$, i.e. for a series-CQD, we have (hereafter, we use $\omega$ to %%@
denote the normalized frequency $\omega/\Gamma$)
\begin{subequations}
\bn
S_R(\omega)&=& S_0+S_1(\omega)+S_2(\omega), \\
S_0 &=& \frac{4 x^2 (8x^4 - 2x^2+1)}{(1+4x^2)^3}\Gamma, \label{eq:zero-noise-of-series}\\
S_1(\omega) &=& \frac{2 x^2}{1+4x^2} \frac{\omega^2}{\omega^2+1}\Gamma,\\
S_2(\omega) &=& - \frac{\omega^2 [(1- 12x^2)\omega^2 +80 x^4 -40x^2 +1]}{[(\omega+2x)^2+1] %%@
[(\omega- 2x)^2+1]} \cr
&& \times \frac{2 x^2}{(1+4x^2)^3} \Gamma, \label{s2}
\en
and
\bq
S_Q(\omega)= \frac{8x^2}{(1+4x^2) (\omega^2+1)} \Gamma.
\eq   
\end{subequations}
It can be obtained from Eq.~(\ref{s2}) that $S(\omega)$ has a peak at $\omega=2x$ if $x<1$, but a %%@
dip otherwise [see Fig.~1(a) below].
If $\Gamma'=\Gamma$ ($\gamma=1$), i.e. a CQD in a completely symmetrical parallel arrangement, we %%@
have
\begin{subequations}
\bn
S_R(\omega)&=& S_0+S_1(\omega), \label{eq:sr-parallel-CQDU0-0} \\
S_0&=& \Gamma, \\
S_1(\omega)&=& \frac{\omega^2 }{\omega^2 + 16} \Gamma, \\
S_Q(\omega)&=& \frac{4}{\omega^2+16} \Gamma, \label{eq:sq-parallel-CQDU0-0}
\en
\end{subequations}
for $\varphi=0$; and for $\varphi=\pm \pi$ we obtain
\begin{subequations}
\bn
S_R(\omega)&=& S_0 +S_1(\omega), \\
S_0 &=& \frac{x^2 (x^4 -x^2 +2)}{(1+x^2)^3} \Gamma, \\
S_1(\omega) &=& x^4 \omega^2 [(5+x^2) \omega^4 + (60-28x^2 - 8x^4) \omega^2
+ 16x^6 \cr
&& +64x^4+16x^2 +224]\Gamma (1+x^2)^{-3} (\omega^2+4)^{-1} \cr
&& \times \{[(\omega-2x)^2+4] [(\omega+2x)^2 + 4]\}^{-1}, \\
S_Q(\omega)&=& \frac{4x^2}{(1+x^2)(\omega^2 +4)}\Gamma.
\en
\end{subequations}
The frequency-dependent part of the noise is thus
\bn
S(\omega)-S_0 &=& \frac{\omega^2 [(3x^2-1) \omega^2 -20x^2 (x^2 -2)
- 4 ]}{[(\omega-2x)^2+4] [(\omega+2x)^2 + 4]} \cr
&& \times \frac{x^2}{(1+x^2)^{3}} \Gamma ,
\en
which indicates that the noise spectrum exhibits an unambiguous dip only when $x\gtrsim 2$ (this %%@
is different from the case of series-CQD revealed above).

\section{Coulomb repulsion effect on finite-frequency shot noise}

This section is concerned with the effects of strong Coulomb repulsion ($U=\infty$) on the noise %%@
spectrum in the case of an extremely large bias-voltage and zero temperature.
In fact, the time-oscillatory current is no longer spatially translationally invariant at finite %%@
frequency due to a time dependent charge accumulation in the device. Coulomb interactions among %%@
electrons screen this accumulation and therefore can significantly influence the noise %%@
spectrum.\cite{Hanke,Korotkov}

In this case, the probability for a double-occupation state completely vanishes, $\rho_{dd}=0$, %%@
and the corresponding number-resolved QREs are:
\begin{subequations}
\label{rateqU} 
\bn
\dot{\rho}_{11}^{(n)} &=& \Gamma \rho_{00}^{(n)} - \Gamma' \rho_{11}^{(n)} +i \Omega ( %%@
\rho_{12}^{(n)}-\rho_{21}^{(n)}) \cr
&& -\frac{1}{2} \sqrt{\Gamma\Gamma'} (e^{-i\varphi/2} \rho_{12}^{(n)} + e^{i\varphi/2} %%@
\rho_{21}^{(n)}), \label{rc11} \\
\dot{\rho}_{22}^{(n)} &=& \Gamma' \rho_{00}^{(n)} - \Gamma \rho_{22}^{(n)} -i\Omega %%@
(\rho_{12}^{(n)}-\rho_{21}^{(n)}) \cr
&& -\frac{1}{2} \sqrt{\Gamma\Gamma'} (e^{-i\varphi/2} \rho_{12}^{(n)} + e^{i\varphi/2} %%@
\rho_{21}^{(n)}), \label{rc21} \\
\dot{\rho}_{12}^{(n)} &=& i\Omega (\rho_{11}^{(n)}-\rho_{22}^{(n)}) - \frac{1}{2} (\Gamma + %%@
\Gamma') \rho_{12}^{(n)} \cr
&& \hspace{-1cm} + \sqrt{\Gamma\Gamma'} e^{-i\varphi/2} \rho_{00}^{(n)} - \frac{1}{2} %%@
\sqrt{\Gamma\Gamma'} e^{i\varphi/2} (\rho_{11}^{(n)} + \rho_{22}^{(n)}),\label{rc31} 
\en
\end{subequations}
together with the normalization relation
$\rho_{00}+\rho_{11}+\rho_{22}=1$. The equation of motion for $\rho_{00}^{(n)}$ is the same as %%@
Eq.~(\ref{nrc01}).

Following the same procedures as indicated in the preceding section, we can evaluate the %%@
stationary current 
\bq
I = \frac{4(\gamma +1)x^2}{(\gamma +1)^2+12 x^2} \Gamma, \label{eq:current-of-series-U}
\eq
and shot noise spectrum for the case of strong Coulomb blockade. 
Straightforwardly, we again obtain analytical expressions for frequency-dependent noise in %%@
several special cases. For a series-CQD, we have
\begin{subequations}
\bn
S_R(\omega)&=& S_0+S_1(\omega), \\
S_0 &=& \frac{8 x^2 (80x^4 - 8x^2+1)}{(1+12x^2)^3}\Gamma, \label{eq:zero-noise-of-series-U} \\
S_1(\omega) &=& \frac{128 x^4 \omega^2}{(1+12x^2)^3} \Gamma [(8+ 16x^2)\omega^4 \cr
&& +(18 -28 x^2 -128x^4)\omega^2 +256x^6 +48 x^4 \cr
&& -64x^2 +13] [\omega^2 (2\omega^2 -8x^2-4)^2 \cr
&& + (1+12x^2 -5\omega^2)^2]^{-1}, \\ 
S_Q(\omega)&=& \frac{16 x^2 }{(1+12x^2)} \Gamma [4 \omega^4 -(5 +32 x^2) \omega^2 +64x^4 \cr
&& +28 x^2 +3] [\omega^2 (2\omega^2 -8x^2-4)^2 \cr
&& + (1+12x^2 -5\omega^2)^2]^{-1}.
\en   
\end{subequations}
In the case of a symmetric parallel-CQD, $\Gamma'=\Gamma$, the noise spectrum for $\varphi=\pm %%@
\pi$ are given by
\begin{subequations}
\bn
S_R(\omega)&=& S_0 +S_1(\omega), \\
S_0 &=& \frac{4 x^2 (5 x^4 -2x^2 +1)}{(1+ 3x^2)^3} \Gamma, \\
S_1(\omega) &=& \frac{16x^4 \omega^2}{(1+3x^2)^{3}} \Gamma [(2+x^2) \omega^4 + (18-7x^2 - 8x^4) %%@
\omega^2 \cr
&& + 16x^6+12x^4-64x^2 +52][\omega^2 (\omega^2 -4x^2-8)^2 \cr
&& + (12x^2+4-5 \omega^2)^2]^{-1}, \\
S_Q(\omega)&=& \frac{4 }{(1+3x^2)} \Gamma [(1+2x^2) \omega^4 - (3-6x^2 + 16x^4) \omega^2 \cr
&& + 32x^6+8x^4-4x^2 -4] [\omega^2 (\omega^2 -4x^2-8)^2 \cr
&& + (12x^2+4-5 \omega^2)^2]^{-1}.
\en
\end{subequations}
In other cases, it is necessary to resort to numerical evaluations (section V).

\section{Numerical calculations and discussion}

In this section, we discuss our numerical calculations of the shot noise power spectrum, for both %%@
the noninteracting ($U=0$) and strongly interacting ($U=\infty$) cases.

\begin{figure}[htb]
\includegraphics[height=4cm,width=8.7cm]{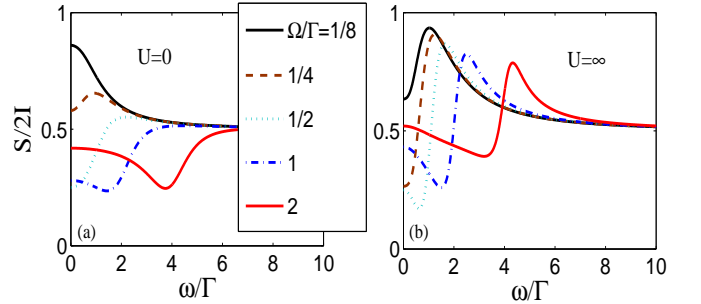}
\caption{(Color online) Fano factor, $F=S(\omega)/2I$, vs. frequency, $\omega/\Gamma$, in a %%@
series-CQD for several values of dot-dot hopping strength, $\Omega/\Gamma=1/8$, $1/4$, $1/2$, %%@
$1$, and $2$, in the cases of absence of interdot Coulomb repulsion $U=0$ (a) and of strong %%@
Coulomb repulsion $U=\infty$ (b).}
\label{fig:snff-series1}
\end{figure}

\begin{figure}[htb]
\includegraphics[height=4cm,width=8.7cm]{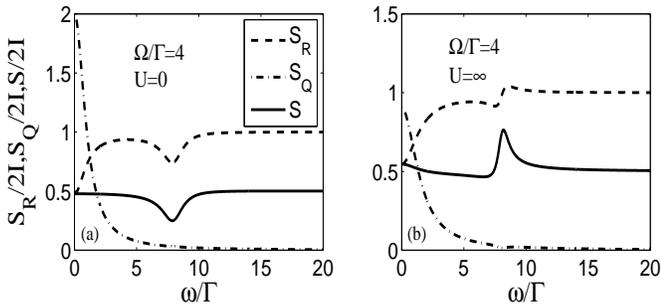}
\caption{Noise spectrum for the right lead, $S_R(\omega)/2I$, the charge correlation spectrum, %%@
$S_Q(\omega)/2I$, and the Fano factor, $F=S(\omega)/2I$, vs. frequency, $\omega/\Gamma$, for a %%@
series-CQD with $\Omega/\Gamma=4$ in the cases of absence of interdot Coulomb repulsion $U=0$ (a) %%@
and of strong Coulomb repulsion $U=\infty$ (b).}
\label{fig:snff-series}
\end{figure}

To start, we examine the noise spectrum of a series-CQD as it depends on increasing dot-dot %%@
hopping strength. The frequency-dependent Fano factor, $F=S(\omega)/2I$, is plotted as a function %%@
of normalized frequency $\omega/\Gamma$ in Figs.~\ref{fig:snff-series1} and %%@
\ref{fig:snff-series}. 
Two rates are of essential importance in determining the noise spectrum: the rate of electrons %%@
entering into the system from the electrodes or its escape rate from the system, $\Gamma$; and %%@
the rate of coherent hopping between the two QDs, because electrons inside the system shuttle %%@
between the two QDs at a frequency $\Omega$. 
In the case of low hopping strength, $\Omega/\Gamma<1$, when an electron from the left lead is %%@
injected into the first QD, no further electrons can enter this QD until this electron transfers %%@
to the second QD with a quite long time scale, $\Omega^{-1}$, for this removal process. Thus, in %%@
this slow hopping limit, increasing $\Omega$ can enhance the transmission probability of an %%@
electron entering into the second dot, and thus reduce the zero-frequency shot noise %%@
(Fig.~\ref{fig:snff-series1}). As the frequency $\omega$ increases, the electron inside the first %%@
QD has a greater opportunity to instantly tunnel into the second QD, which enhances the shot %%@
noise. Consequently, the noise spectrum exhibits a nonzero-frequency peak as shown in %%@
Fig.~\ref{fig:snff-series1}.  
While, for high values of hopping strength, $\Omega/\Gamma>1$, the electron has a relatively high %%@
probability to enter into the second QD (once it is injected into the first QD from the left %%@
lead), it is also true that the electron also has high probability to return {\em back} to the %%@
first QD from the second QD periodically at a frequency $2\Omega$ (the Rabi frequency), before it %%@
escapes to the right lead, which takes place on a relatively long time scale determined by %%@
$\Gamma^{-1}$. The rapid return of the electron to the first dot blocks the entry of another %%@
electron into the dot, leading to a suppression of noise at $\omega=2\Omega$ in the absence of %%@
interdot Coulomb repulsion, as shown in Fig.~\ref{fig:snff-series1}(a) and %%@
\ref{fig:snff-series}(a). Similar noise properties were found in previous studies for a coupled %%@
double well structure\cite{Sun} and for a CQD.\cite{Djuric1} 
  
Interestingly, we find that strong interdot Coulomb repulsion also plays a crucial role in %%@
determining noise. In Figs.~\ref{fig:snff-series1}(b) and \ref{fig:snff-series}(b), the noise %%@
spectrum exhibits a peak at $\omega=2\Omega$ for the system with strong dot-dot hopping and %%@
infinite Coulomb repulsion, instead of a dip for the system without interdot Coulomb repulsion. A %%@
similar Coulomb interaction effect on the noise spectrum was also reported for a single QD %%@
connected to two ferromagnetic electrodes.\cite{Gurvitz3} For the sake of comparison, we also %%@
plot the noise spectrum for the leads, $S_R(\omega)/2I$, and the charge correlation noise %%@
spectrum, $S_Q(\omega)/2I$, in Fig.~\ref{fig:snff-series}. It is evident that Coulomb repulsion %%@
mainly influences noise in the leads, and it reduces the charge correlation spectrum by nearly a %%@
factor of $2$, but has little effect on its shape. 

Moreover, strong dot-dot hopping strength, $\Omega$, combines the two dots rather tightly, and %%@
thus the CQD behaves much more like a single QD. As a result, the zero-frequency shot noise %%@
reduces to the result of a symmetric single QD, $F_0=1/2$ in the limit of %%@
$\Omega/\Gamma\rightarrow\infty$. This can be easily verified from %%@
Eqs.~(\ref{eq:current-of-series}) and (\ref{eq:zero-noise-of-series}) for $x\rightarrow\infty$. %%@
Strong Coulomb repulsion results in a bit larger Fano factor, $F_0=5/9$, from %%@
Eqs.~(\ref{eq:current-of-series-U}) and (\ref{eq:zero-noise-of-series-U}). 

\begin{figure}[htb]
\includegraphics[height=4cm,width=8.7cm]{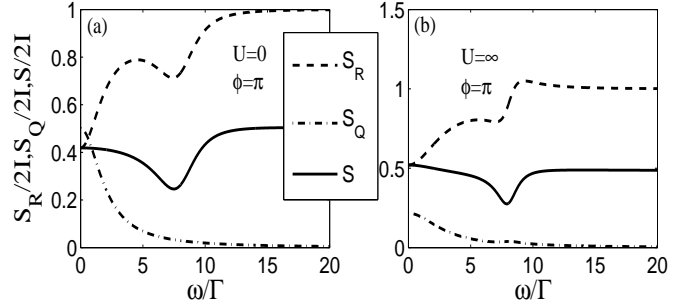}
\caption{Noise spectrum for the right lead, $S_R(\omega)/2I$, the charge correlation noise %%@
spectrum, $S_Q(\omega)/2I$, and the Fano factor, $F=S(\omega)/2I$, vs. frequency, %%@
$\omega/\Gamma$, for a parallel-CQD with $\Omega/\Gamma=4$, $\Gamma'/\Gamma=1$ and a %%@
magnetic-flux $\varphi=\pi$, in the cases of absence of interdot Coulomb repulsion $U=0$ (a) and %%@
of strong Coulomb repulsion $U=\infty$ (b).}
\label{fig:snff-parallel}
\end{figure}

We have also examined the quantum interference effect between the two distinct pathways of a CQD %%@
in a parallel arrangement on the noise power spectrum. We plot the frequency-dependent noise %%@
spectra for a completely symmetrical CQD interferometer ($\gamma=1$) with $x=\Omega/\Gamma=4$ at %%@
the magnetic-flux $\varphi=\pi$, in Fig.~\ref{fig:snff-parallel}. It is evident from %%@
Fig.~\ref{fig:snff-parallel} that the noise spectrum of a parallel-CQD has similar properties as %%@
that of a series-CQD, i.e., has a dip at the Rabi frequency, in the absence of Coulomb %%@
interaction; but (2) it is different from a series-CQD in the presence of strong interdot Coulomb %%@
repulsion, i.e., it still exhibits a dip at the Rabi frequency.  

\begin{figure}[htb]
\includegraphics[height=6.5cm,width=8.7cm]{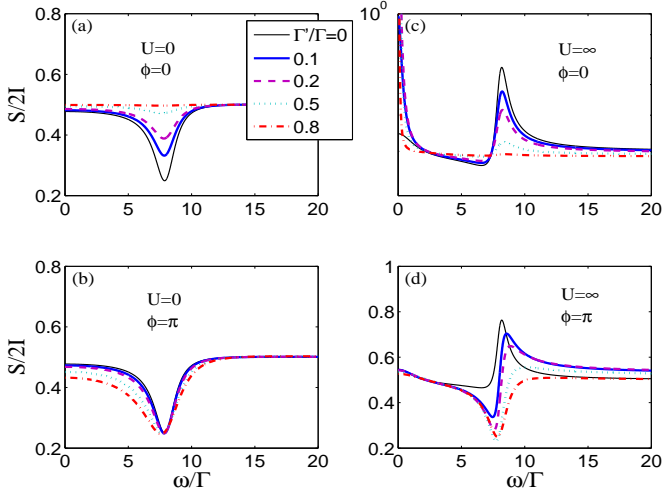}
\caption{(Color online) Fano factor, $F=S(\omega)/2I$, vs. frequency, $\omega/\Gamma$, for a CQD %%@
with $\Omega/\Gamma=4$ in the cases of absence of interdot Coulomb repulsion $U=0$ (a,b) and of %%@
strong Coulomb repulsion $U=\infty$ (c,d), as well as with different magnetic-fluxes, $\varphi=0$ %%@
(a,c) and $\varphi/\pi=1$ (b,d) for several values of additional tunneling strengths, %%@
$\Gamma'/\Gamma=0$, $0.1$, $0.2$, $0.5$, and $0.8$.}
\label{fig:snffg}
\end{figure}

To explore the quantum interference effect more fully, we analyze the noise spectrum as it %%@
depends on increasing tunneling strength, $\Gamma'$, of the additional pathway. Figure %%@
\ref{fig:snffg} exhibits our numerical results for noise spectra with magnetic-fluxes, %%@
$\varphi=0$ and $\pi$, including the cases of no Coulomb repulsion and of strong Coulomb %%@
repulsion. It is clear that (1) increase of tunneling strength of the additional path branch %%@
continuously decreases the dip of the noise spectrum at $\omega=2\Omega$ if $\varphi=0$, while it %%@
continuously widens the dip if $\varphi=\pi$ for a system with no Coulomb repulsion %%@
[Figs.~\ref{fig:snffg}(a) and (b)]; (2) the combined effect of strong Coulomb repulsion and full %%@
interference gives rise to substantially enhanced zero-frequency noise, as pointed out previously %%@
in the absence of magnetic-flux,\cite{Dong4} but it decreases very rapidly to sub-Poissonian %%@
values at finite frequencies, and also its peak at $\omega=2\Omega$ is gradually suppressed by %%@
the interference effect [Fig.~\ref{fig:snffg}(c)]; and (3) the noise spectrum exhibits a %%@
transition peak--Fano-type structure--dip with increasing tunneling strength of the additional %%@
path branch for $\varphi=\pi$ [Fig.~\ref{fig:snffg}(d)].         

\begin{figure}[htb]
\includegraphics[height=8cm,width=8cm]{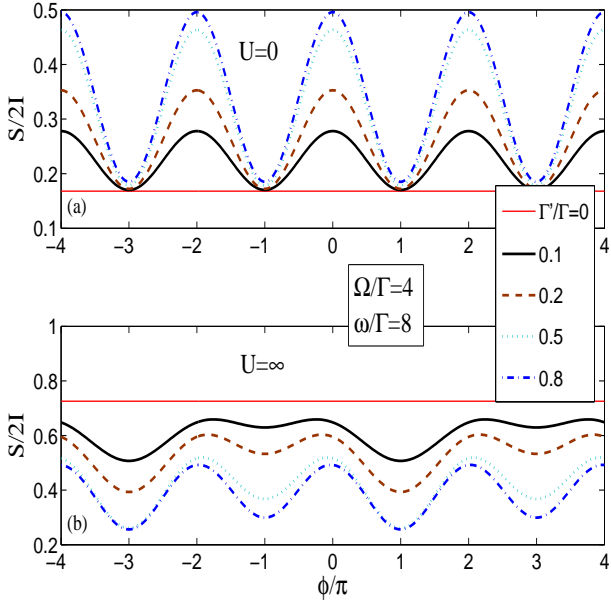}
\caption{(Color online) Calculated Fano factor, $F=S(\omega)/2I$ at the Rabi frequency %%@
$\omega/\Gamma=8$, vs. magnetic-flux, $\varphi/\pi$, for various values of $\Gamma'/\Gamma$ with %%@
$\Omega/\Gamma=4$ in the cases of absence of interdot Coulomb repulsion, $U=0$ (a), and of strong %%@
Coulomb repulsion, $U=\infty$ (b).}
\label{fig:snffvp}
\end{figure}

Finally, we exhibit the magnetic-flux dependence of the noise spectrum at $\omega=2\Omega$ %%@
(Fig.~\ref{fig:snffvp}). It is evident that the noise spectrum is a periodic function of %%@
magnetic-flux, and the effect of strong Coulomb repulsion is to vary the period of the noise %%@
spectrum from $2\pi$ to $4\pi$.

\section{Conclusions}

In summary, we have analyzed the frequency-dependent current noise for coherent resonant %%@
tunneling through a CQD-AB interferometer in the case of an extremely high bias-voltage and zero %%@
temperature by means of number-resolved QREs and MacDonald's formula. Our attention has been %%@
focused on the role of coherent dot-dot hopping as well as on the interference effect between two %%@
distinct path branches involving magnetic-flux dependence, and also on inter-dot Coulomb %%@
repulsion effects.  
We have derived explicit analytical expressions for noise power spectrum of a series-CQD and of a %%@
completely symmetric parallel-CQD for specific magnetic-fluxes, $\varphi=\pi$, in both cases of %%@
no inter-dot Coulomb interaction and of infinite inter-dot Coulomb repulsion. Our numerical %%@
calculations have shown that (1) in absence of Coulomb interaction, the noise spectrum of a %%@
series-CQD has a peak at the Rabi frequency for weaker dot-dot hopping, $\Omega/\Gamma<1$, but %%@
becomes an unambiguous dip for stronger dot-dot hopping, $\Omega/\Gamma \gtrsim 1$; (2) it always %%@
exhibits a peak due to Coulomb repulsion; and (3) this peak can gradually become a dip again for %%@
a parallel-CQD due to variation of interference pattern by tuning enclosed magnetic-flux.

\begin{acknowledgments} 

This work was supported by Projects of the National Science Foundation of China, the Shanghai %%@
Municipal Commission of Science and Technology, the Shanghai Pujiang Program, and Program for New %%@
Century Excellent Talents in University (NCET).

\end{acknowledgments}

\end{document}